**Topotactic phase transformation of the brownmillerite $SrCoO_{2.5}$ to the perovskite $SrCoO_{3-\delta}$**

Hyoungjeen Jeen[1], Woo Seok Choi[1], John W. Freeland[2], Hiromichi Ohta[3], Chang Uk Jung[1,4], and Ho Nyung Lee[1*]

[1]Materials Science and Technology Division, Oak Ridge National Laboratory, Oak Ridge, TN, 37831, USA
[2]Advanced Photon Source, Argonne National Laboratory, Argonne, Illinois 60439, USA.
[3]Research Institute for Electronic Science, Hokkaido University, Sapporo 001-0020, Japan
[4]Department of Physics, Hankuk University of Foreign Studies, Yongin, 449-791, South Korea

*E-mail: hnlee@ornl.gov

Oxygen stoichiometry is one of the most important elements in determining the physical properties of transition metal oxides (TMOs). A small change in the oxygen content results in the variation of valence state of the transition metal, drastically modifying the materials functionalities. The latter includes, for instance, (super-)conductivity, magnetism, ferroelectricity, bulk ionic conduction, and catalytic surface reactions.[1-5] In particular, among those applications, TMOs with mixed valence states have attracted attention for many environmental and renewable energy applications, including catalysts, hydrogen generation from water splitting, cathodes in rechargeable batteries and solid oxide fuel cells, and oxygen separation membranes.[6-8] For example, previous studies have shown that the ability to control the number of $d$-band electron population and detailed spin configuration in TMOs is critical for improved catalytic performance of TMOs.[9, 10] In this context, $SrCoO_x$ ($2.5 \leq x \leq 3.0$) is an ideal class of materials to study the evolution of the physical properties by modifying the valence state in TMOs, due to the existence of two structurally distinct topotactic phases, i.e. the brownmillerite $SrCoO_{2.5}$ (BM-SCO) (see **Figure 1**a) and the perovskite $SrCoO_3$.[11, 12] Especially, BM-SCO has atomically-ordered one-dimensional vacancy channels (see Figure 1a), which can accommodate additional oxygen when the valence state of Co is changed. Moreover, $SrCoO_x$ exhibits a wide spectrum of physical properties from antiferromagnetic insulator to ferromagnetic metal depending on the oxygen stoichiometry.[11-13] Since $SrCoO_x$ has only a single control knob, i.e. the oxygen content $x$, to modify the Co valence state without cation doping, it is straightforward to study the valence state (i.e. oxygen content) dependent physical properties. However, so far, the growth of high quality single crystalline materials has not been as successful due to difficulty in controlling the right oxidation state.

In this work, we report on the epitaxial growth of high quality BM-SCO single crystalline films on $SrTiO_3$ (STO) substrates by pulsed laser epitaxy (PLE). In order to examine the topotactic phase transformation to the perovskite $SrCoO_{3-\delta}$ (P-SCO), some of the samples were subsequently *in-situ* annealed at various oxygen pressure ($P(O_2)$) to fill the oxygen vacancies. While the direct growth of P-SCO films with $x = 3.0$ was an arduous task, we found that post-annealing in high $P(O_2)$ (> several hundreds of Torr) could fill sufficient amount of oxygen vacancies, yielding systematic evolution in electronic, magnetic, and thermoelectric properties.

Figure 1b shows an x-ray diffraction (XRD) $\theta–2\theta$ scan pattern of a BM-SCO film (25 nm in thickness). Note that BM-SCO is orthorhombic with lattice constants of $a_o = 5.5739$, $b_o = 5.4697$, and $c_o = 15.7450$ Å,[14] which can be represented as pseudo-tetragonal ($a_t = 3.905$ and $c_t/2 = 3.936$ Å) to estimate the lattice mismatch. The orthorhombic notation will be used throughout this paper. All the films were grown on (001) STO substrates ($a = 3.905$ Å). By comparing the lattice mismatches along different crystallographic directions, it is clear that the BM-SCO phase prefers to grow as *c*-axis-oriented on STO substrates due to the nearly zero in-plane mismatch, as the XRD scan shown in Figure 1b clearly confirms. The characteristic increase of the $c$ lattice constant ($c/2$ = 7.860 Å), originating from the alternatively-stacked octahedral and tetrahedral sub-layers, is clearly seen. X-ray reciprocal space mapping (RSM) shown in **Figure 2**a indeed confirms the coherent growth on the perfectly lattice matched STO substrate, and sharp peaks with clear Kiessig fringes (Figure 1b) demonstrate that the films are of high quality. X-ray rocking curve $\omega$-scans revealed a full width half maximum (FWHM) of < 0.04°, demonstrating the excellent crystallinity (*cf.*, FWHM of the 002 STO peak was ~0.02°) of our films (data not shown).

While we have shown the XRD data from a well-optimized, high quality thin film, it is worthwhile to mention that the BM-SCO thin film could only be synthesized in a very narrow growth window, as we have learned through a systematic growth of several tens of samples as shown in Figure 2d. This is most likely related to the multiple oxidization states offered by Co. Note that the two valence states, $Co^{2+}$ ($3d^7$) and $Co^{3+}$ ($3d^6$), are most commonly observed ones. In particular, CoO impurity phases were readily found in XRD $\theta$-$2\theta$ scans, when films were grown at low $P(O_2)$ (< 10 mTorr). The impurity phases could be avoided by increasing the $P(O_2)$, up to 200 mTorr. We also found the film's phase purity was sensitive to the growth temperature, which plays a significant role in preparing films with the right stoichiometry. We found 750 °C as the optimal growth temperature on STO. When thin films were grown at lower temperatures, we often observed the inclusion of impurity phases, most likely rhombohedral $Sr_6Co_5O_{15-\delta}$, in addition to the BM phase. This can be explained by the easy structural conversion from the BM structure to a low temperature polymorph phase, e.g. $Sr_6Co_5O_{15-\delta}$,[15, 16] which is thermodynamically favored at low growth temperatures and low $P(O_2)$. We note that the pressure for the post-growth cooling was kept the same as the growth $P(O_2)$ to minimize the oxygen content fluctuation, since the oxygen content in cobaltite thin films could change even during the post-growth cool down.

Although we were able to optimize the growth condition for high quality BM-SCO films, the effort to directly grow thin films with higher oxygen content, i.e. $x > 2.5$ in $SrCoO_x$, was unsuccessful. As mentioned before, the growth in extremely high $P(O_2)$ (for instance, even in several hundreds of mTorr) could not stabilize the perovskite phase, but rather yielded Co-O impurity phases. Note that most perovskite oxides can be readily grown under such high $P(O_2)$ without oxygen vacancies.[17-19] This might be related to the fact that the thermodynamic barrier for the formation of the perovskite phase with $Co^{4+}$ is so high that energetically more favorable phases with $Co^{2+}$ and/or $Co^{3+}$ are preferentially formed under typical PLE conditions. It is worth mentioning that, in case of other perovskites, for example, manganite thin films, the formation of perovskite phase is always preferred over the brownmillerite one. In case of manganite thin films, the growth even in high vacuum yields only perovskite films.[20] Interestingly, the use of an oxygen getter layer next to the manganite film was found to offer a way to stabilizing the Mn-based brownmillerite phase.[21]

Since the direct growth of $SrCoO_x$ films with $x > 2.5$ was not possible as also confirmed by previous attempts,[22] we tried *in-situ* oxygen postannealing in much higher $P(O_2)$ to trigger the topotactic oxidation to form SCO phases with x>2.5. Figure 1c, d show the XRD $\theta$-$2\theta$ scan results from samples postannealed at 600 °C for 5 minutes in $P(O_2)$ = 300 and 600 Torr, respectively. Their corresponding RSMs are shown in Figure 2b, c. As shown in Figure 1d, when the BM-SCO thin films were annealed at 600 °C and at $P(O_2)$ = 600 Torr, the films showed complete suppression of the half-order peaks from the BM phase. This temperature could be as low as 400 °C, but annealing at higher temperatures (> 600 °C) resulted in significant deterioration of the film's structural quality, yielding polycrystalline films with Co-O impurity phases. A new set of XRD peaks revealing a significant decrease in the lattice constant (by ~3%) along the *c*-axis direction was subsequently appeared. The reduction in the *c*-axis lattice constant indicates successful oxygen intercalation into the tetrahedral sub-lattices in the BM phase. The disappearance of the half-order peaks also indicates removal of the chemical and structural contrasts between the octahedral and tetrahedral sub-layers by oxygen intercalation. This conversion, thus, implies that the P-SCO structure has been successfully formed. According to the literature,[23] the complete disappearance of the characteristic XRD peaks from BM-SCO occurs near $x = 2.75$.

Furthermore, the formation of the P-SCO phase was sensitive to the annealing pressure. As shown in Figure 1c and 2b, a mixed-phase of BM- and P-SCO (Mix-SCO) was dominantly observed when a BM-SCO film was post-annealed at lower $P(O_2)$ ($100 < P(O_2) \leq 300$ Torr). While it was unclear how the two phases are physically positioned within the film, the XRD peak positions of the both phases in the Mix-SCO film were identical to those from individual BM-SCO and P-SCO films. Moreover, we could not find any peak corresponding to a possible intermediate phase, which suggests that the structural conversion from BM- to P-SCO is a first order-like phase transition. The high $P(O_2)$ (>300 Torr) required for converting into P-SCO seems to explain why we could not directly grow P-SCO in $O_2$. (Note that PLE growth in such high background $P(O_2)$ is impractical due to the highly increased scattering of laser-ablated species by the background gas.)

We further note that the filling of oxygen vacancies seemed to subsequently induce a strain relaxation of the film most probably due to the large lattice mismatch (2%) between the P-SCO film and the STO substrate. As comparatively shown in Figure 2a-c, the BM-SCO phase maintained a fully-strained state (Figure 2a, b), while the P-SCO phase was relaxed (Figure 2b, c). Note that the Mix-SCO film shown in Figure 2b still has the fully-strained BM phase.

Although the overall structural evolution with the oxygen intercalation has been confirmed, an accurate determination of the valence state or exact amount oxygen content from the XRD data is challenging. Therefore, we used polarized x-ray absorption spectroscopy (XAS) to probe elementally resolved details of the chemical valence state. Since the formation of oxygen vacancies alters the valence state of Co in SCO, the change in oxygen stoichiometry can be tracked by checking the Co valence state. **Figure 3**a shows XAS data of Co *L*-edge. From P-SCO, a shift of peaks towards the higher energy was observed as compared to those from BM-SCO. The peak energy shift clearly indicates a change in the Co valence state from 3+ toward 4+ with oxygen intercalation.[24, 25] As expected, the overall trend represents that the P-SCO film has a higher valence state of Co ions than that of BM-SCO. We further compared O *K*-edge XAS data for both BM-SCO and P-SCO films. The data consists of different hybridizations between O-2*p* and neighboring cation orbitals, i.e. Co-3*d* and Sr-4*d*. Among them, the pre-peak at ~526.5 eV in the O *K*-edge originates from the hybridization between O-2*p* and Co-3*d*. Note that this pre-peak is highly sensitive to the oxygen content, i.e. the pre-peak intensity increases substantially as the Co valence state changes from 3+ toward 4+.[24] As shown in Figure 3b, the pre-peak is distinctively visible near 527 eV in P-SCO, while it is highly suppressed in case of BM-SCO. This clearly indicates that both oxygen intercalation into BM-SCO and subsequent change in the Co valence state were successful. Note that this XAS was measured in bulk-sensitive fluorescence yield, and features at higher energy are obscured due to substrate contributions.

Changes in the magnetic properties by valence state modification were further confirmed by measuring temperature dependent magnetization, $M(T)$, and magnetic hysteresis curves, $M(H)$ by a superconducting quantum interference device (SQUID) magnetometer. As shown in **Figure 4**a, the magnetization in BM-SCO is nearly zero, and there is no characteristic change in $M(T)$, i.e. magnetic phase transition. In addition, both $M(H)$ measurements performed at 10 K (Figure 4b) and 250 K (data not shown) revealed similar hysteresis curves with zero coercive fields. Since the BM-SCO film was grown with zero lattice mismatch, these magnetic behaviors in our BM-SCO thin film are consistent with the antiferromagnetism found in bulk BM-SCO materials. Note the latter have $T_N$ = 570 K.[11, 26] This bulk-like behavior clearly indicates that our BM-SCO epitaxial thin films are homogeneously grown without any oxygen enriched SCO phases or paramagnetic impurities,[27, 28] consistent with our XRD results. On the other hand, the oxygen post-annealed films showed clear evidence of ferromagnetism (Figure 4a, b).[11] The ferromagnetism originates from the emergence of $Co^{4+}$ by oxygen intercalation, which results in an increase in the ferromagnetic exchange interaction.[13] By assuming that pure P-SCO was stabilized, we could estimate the oxygen content within the P-SCO film from $T_C$ and the saturation magnetization. Note that we do not consider the role of strain for this estimation here, as P-SCO is relaxed. The $T_C$ was in between the bulk values from SCO with $x$ = 2.75 and 2.88.[13, 29] In addition, the saturation magnetization (~0.7 $\mu_B$/Co at 6 T) was larger than that of bulk SCO with $x$ = 2.75. Therefore, we conclude that the maximum oxidation state ($x$) achievable in P-SCO on STO by *in situ* oxygen annealing without deteriorating the epitaxy is 2.75 < $x$ < 2.88. A further study on the strain (or lattice mismatch)-dependent oxidation would give more insight into the role of strain on oxidation and related physical properties.

Moreover, note that $M(T)$ measurements showed similar $T_C$ (~200 K) from both Mix- and P-SCO films, suggesting that the Mix-SCO is indeed a physical mixture of BM and P with mixed valence states in Co ions. It further implies that the phase conversion from BM- to P-SCO or oxygen intercalation is a first-order type transition as mentioned earlier, even though a complete oxidation seems to be challenging. Nevertheless, it is worth mentioning that an epitaxial film with a mixed valence state can be achieved without cation doping. By comparing the remnant magnetization between the phase-pure P-SCO and Mix-SCO ($M_r$ = 47 and 22 emu cm$^{-3}$, respectively), we estimate that the fraction of P-SCO in the Mix-SCO thin film is ~47%.

In order to further investigate the electronic transport properties of SCO, we characterized temperature dependent *dc* transport and thermoelectric properties. As shown in Figure 4c, the BM-SCO thin film showed a highly insulating behavior. Based on the electronic transport data, the thermal activation energy was calculated to be 0.19 eV. The value is similar to the activation energy of bulk SrCoO$_{2.5}$ (0.24 eV).[11] On the other hand, a significant reduction in resistivity (more than three orders of magnitude even at room temperature) in the P-SCO thin film was observed. The temperature dependent behavior was also clearly different from that of the BM-SCO. However, a clear metallic behavior, i.e. the reduction of resistivity by decreasing the temperature was not observed from P-SCO on STO. A similar $\rho(T)$ behavior was also seen from the Mix-SCO, even though the overall resistivity was a bit higher than that of P-SCO as shown in Figure 4c. Based on the similarity of the two $\rho(T)$ curves, we conclude that the insulator-like behavior in P-SCO originates from the lack of long-range order induced by the incomplete oxidation mentioned before.

Although we could not observe a clear metallic behavior from P-SCO on STO, the characteristic structures of our SCO films may offer useful other physical properties. Thus, as shown in Figure 4d, we measured the thermopower of SCO thin films at 300 K. Note that the thermoelectric measurement is particularly sensitive to the change in the carrier concentration from hole doping and/or variation in oxygen content.[30-32] First, a systematic decrease in the thermopower was observed upon increase of oxygen content or hole doping in SCO thin films. This empirically implies that the carrier concentration of SCO is increased when oxygen is intercalated into the films. It also agrees well with the systematically enhanced electronic conductivity as shown in Figure 4c. In addition, the positive thermopower values at room temperature in all samples clearly indicate the *p*-type conduction in SCO films. This is caused by hole doping via oxygen intercalation and is consistent with hole carriers in stoichiometric $SrCoO_3$.[27, 33]

In summary, high quality BM-SCO epitaxial thin films were successfully grown on STO substrates within a narrow growth window by PLE. The BM-SCO single crystalline films grown on lattice matched STO clearly exhibited alternating octahedral and tetrahedral layers. We also found that *in situ* post-annealing under high oxygen pressure (>300 Torr of $O_2$) exhibited a clear phase transition from the brownmillerite $SrCoO_{2.5}$ to the perovskite $SrCoO_{3-\delta}$ ($\delta > 0.12$). A mixed valence state was also observed where a BM-SCO was annealed in $100 < P(O_2) \leq 300$ Torr of $P(O_2)$. Overall, the phase transition, i.e. filling of oxygen vacancy channels, substantially changed the physical properties, including electronic transport, magnetic ground states, and electronic structures. Therefore, based our demonstration of successful epitaxial synthesis and topotactic phase control, further studies on electrochemical or surface catalytic effects with the two topotactic phases films with the distinctly different oxygen concentrations and Co valence states may offer an opportunity to develop high performance perovskite-based oxide ionic devices.

EXPERIMENTAL SECTION.

The SrCoO$_x$ (SCO) thin films were grown on STO substrates using PLE (KrF, $\lambda$ = 248 nm). The substrates were etched and thermally treated for achieving TiO$_2$ terminated surfaces. Growth of BM-SCO films was conducted at 650 ~ 800 °C and 0.1 ~ 500 mTorr of oxygen. The laser fluence and repetition rate were fixed at 1.7 J cm$^{-2}$ and 5 Hz, respectively. For *in situ* oxygen post-annealing, as grown BM-SCO thin films were cooled to 600 °C and annealed for 5 min under various oxygen pressure. The samples were structurally characterized with high resolution four-circle XRD (X'Pert, Panalytical Inc.). The details of the valence states in SCO were characterized by XAS at beamline 4-ID-C of the Advanced Photon Source, Argonne National Laboratory using both surface-sensitive electron yield and bulk-sensitive fluorescence yield. Magnetic property was measured with a 7 T SQUID (Quantum Design). Temperature dependent *dc* transport measurements were performed with a 14 T physical property measurement system (Quantum Design). The thermopower values were measured by a conventional steady state method using two Peltier devices under the thin films to give a temperature difference ($\Delta V$~10 K).

ACKNOWLEDGEMENTS.
This work was supported by the U.S. Department of Energy, Basic Energy Sciences, Material Sciences and Engineering Division. Use of the Advanced Photon Source was supported by the U. S. Department of Energy, Office of Science, under Contract No.DE-AC02-06CH11357. H. O. was supported by MEXT (22360271).

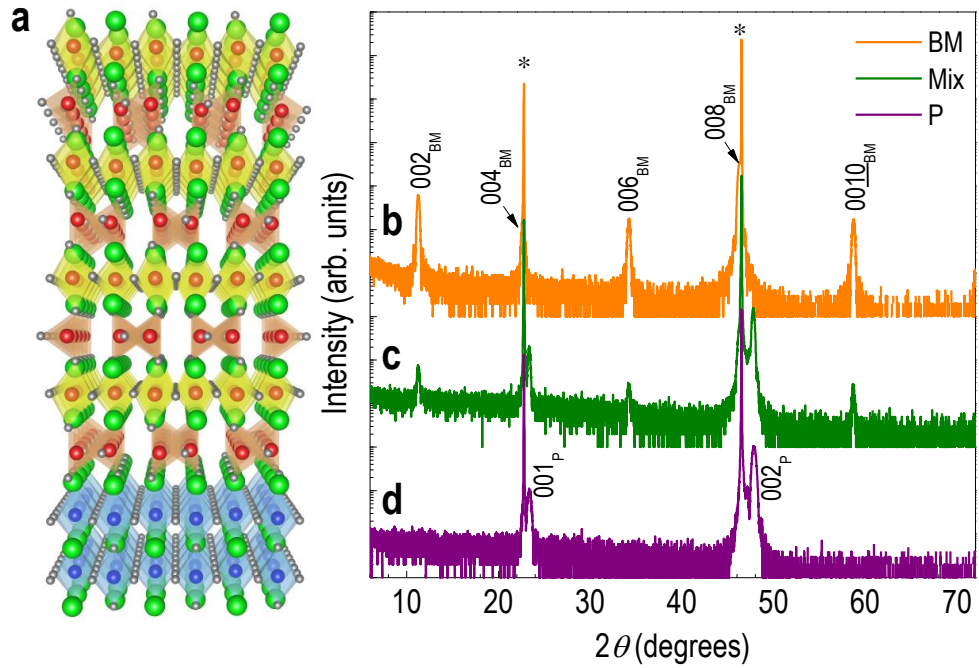

**Figure 1.** a) Schematic of a *c*-axis-oriented BM-SCO film epitaxially grown on a (001) STO substrate. XRD $\theta$–$2\theta$ patterns of b) BM-, c) Mix-, and d) P-SCO thin films on STO substrates. Note that the Mix-SCO and P-SCO were *in situ* annealed at 600 °C, respectively, at 300 and 600 Torr of $P(O_2)$ for 5 min. STO substrate peaks are indicated with asterisks (*).

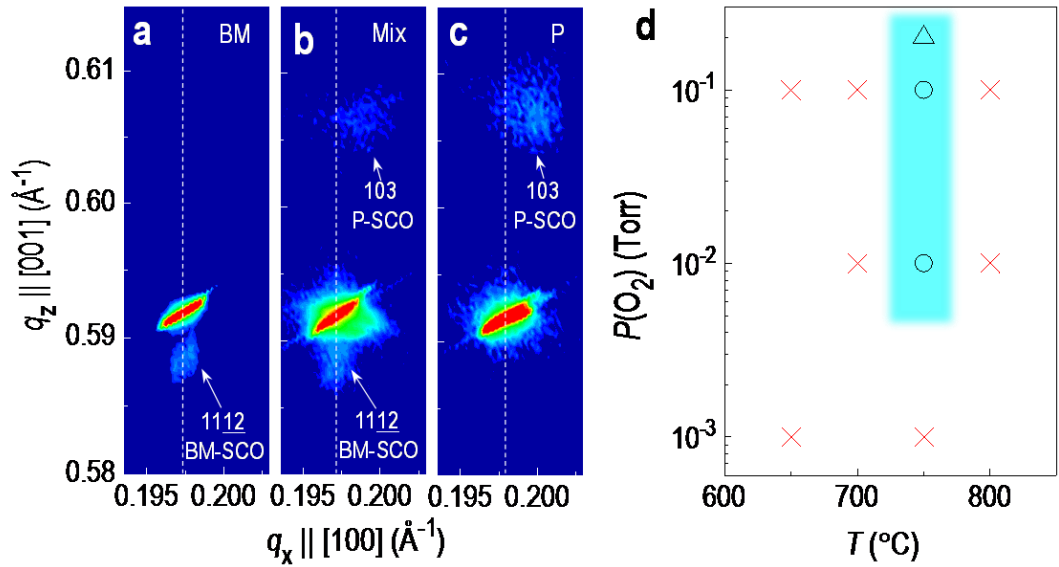

**Figure 2.** Reciprocal space maps from a) BM-, b) Mix-, and c) P-SCO films around the 103 STO Bragg reflection. d) Growth phase diagram for epitaxial BM-SCO films as a function of the oxygen partial pressure and growth temperature. Phase-pure BM-SCO epitaxial films without impurity phases (mainly Co-O) could only be grown in a narrow growth window (the cyan-colored region with open circle and triangle symbols). When the growth was performed under the conditions marked with cross symbols, the resulting films were not epitaxial or contained impurity phases. The open triangle indicates a film grown with poor crystallinity, even though no impurity phases were observed.

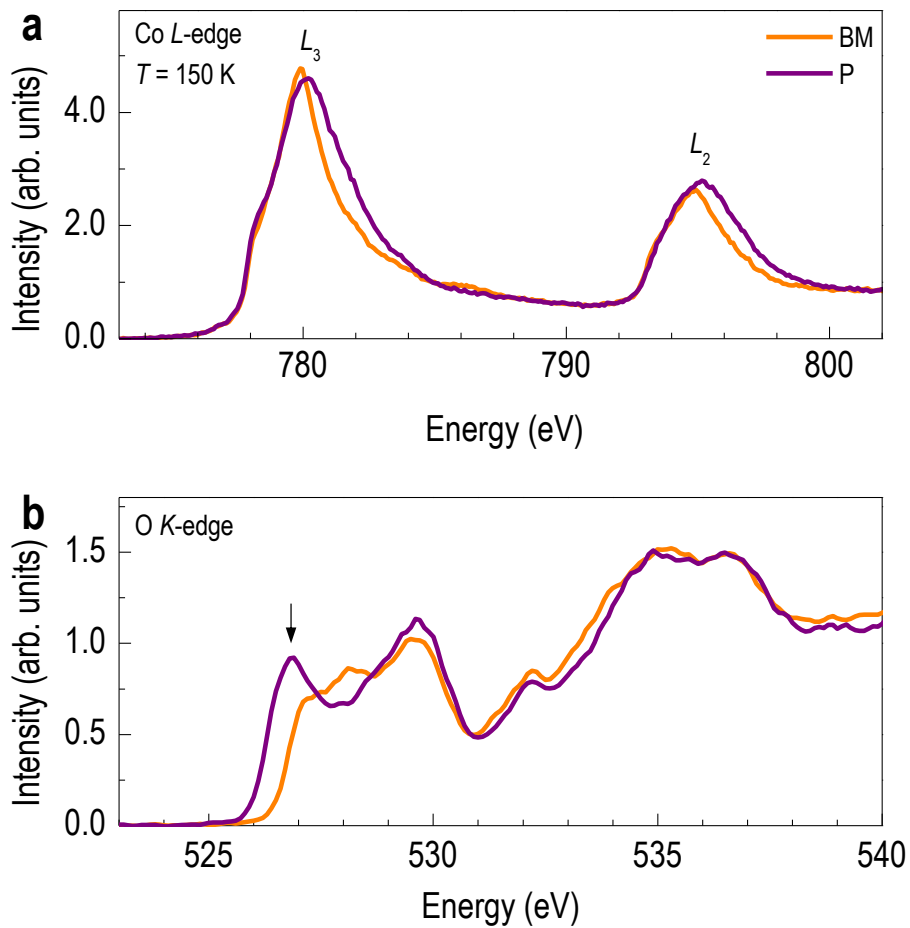

**Figure 3.** a) XAS Co *L*-edge spectra of BM- and P-SCO thin films showing a peak shift towards the higher energy upon oxidation, confirming the Co valence change. b) XAS O *K*-edge spectra of the corresponding samples. A clear development of the oxygen pre-peak (see the arrow) is observed from the P-SCO film upon oxygen intercalation.

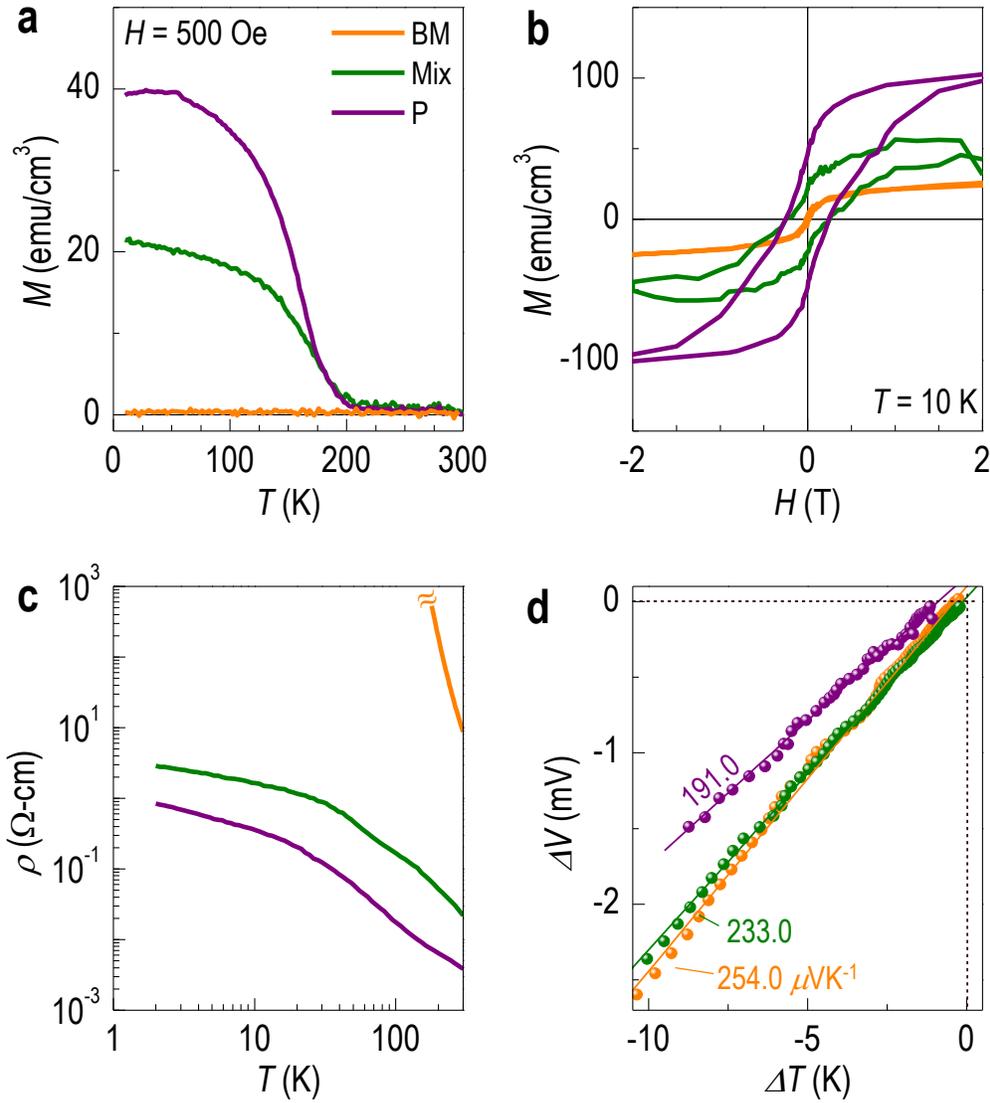

**Figure 4.** a) Temperature dependent magnetization of SCO thin films. While the P-SCO film shows a higher magnetization value than that of the Mix-SCO film, both films reveal the same $T_c$ (~200 K). This confirms that the Mix-SCO film contains a mixture of BM- and P-phases, which were indicted from the XRD results. b) In-plane magnetization hysteresis loops of the three SCO thin films at 10 K. Note that, for the P-SCO sample, 100 emu cm$^{-3}$ is corresponding to 0.62 $\mu_B$/Co. c) Temperature dependent resistivity data, showing a clear reduction of resistivity upon oxidation. d) Thermoelectromotive force of SCO thin films at 300 K. The positive thermopower values confirm the *p*-type conduction in our SCO films.